\documentclass[a4paper,9pt]{article}

\usepackage{hyperref}
\usepackage{amsmath,amsthm,amssymb}
\usepackage{xcolor}
\usepackage{graphicx}
\usepackage{epstopdf}
\usepackage{overpic}
\usepackage{bbold}
\usepackage{textcomp}
\usepackage{siunitx}
\usepackage{physics}
\usepackage{subcaption}
\usepackage{mathtools}
\usepackage[frozencache,cachedir=minted_cache]{minted}
\setminted{fontsize=\footnotesize}
\definecolor{bg}{rgb}{0.95,0.95,0.95}
\newenvironment{code}{\captionsetup{type=listing}}{}

\usepackage{authblk}

\renewcommand{\vec}[1]{\boldsymbol{#1}}
\newcommand{\dx}{\,\text{d}\vec{x}}
\newcommand{\dA}{\,\text{d}\vec{A}}

\newcommand{\jump}[1]{\ensuremath{[\![#1]\!]} }

\DeclareMathOperator{\sign}{sign}

\begin{document}

\title{magnum.np -- A PyTorch based GPU enhanced Finite Difference Micromagnetic Simulation Framework for High Level Development and Inverse Design}

\author[1]{Florian Bruckner \thanks{florian.bruckner@univie.ac.at}}
\author[1]{Sabri Koraltan}
\author[1]{Claas Abert}
\author[1]{Dieter Suess}

\affil[1]{Faculty of Physics, University of Vienna, Austria}
\maketitle

\begin{abstract}
magnum.np is a micromagnetic finite-difference library completely based on the tensor library PyTorch. The use of such a high level library leads to a highly maintainable and extensible code base which is the ideal candidate for the investigation of novel algorithms and modeling approaches. On the other hand magnum.np benefits from the devices abstraction and optimizations of PyTorch enabling the efficient execution of micromagnetic simulations on a number of computational platforms including GPU and potentially TPU systems.
We demonstrate a competitive performance to state-of-the art micromagnetic codes such a mumax3 and show how our code enables the rapid implementation of new functionality. Furthermore, handling inverse problems becomes possible by using PyTorch's autograd feature.
\end{abstract}

\section{Introduction}
Micromagnetic simulations are widely used in a range of applications, from magnetic storage technologies and the design of hard and soft magnetic materials, to the modern fields of magnonics, spintronics, or even neuromorphic computing.
A finite difference approximation has been proven useful for many applications due to its simplicity and its high performance, compared with the more flexible finite element approach.

Currently, there are already many open-source finite difference codes available, like OOMMF \cite{donahue1999oommf}, mumax3 \cite{vansteenkiste2014design}, magnum.af \cite{heistracher2020hybrid}, magnum.fd \cite{abert2014magnumfd}, fidimag \cite{bisotti2020fidimag}, to mention just a few.  However, for the development of new algorithms or for bleeding edge simulations one often needs to modify or extend the provided tools. For example post-processing of the created data often requires the setup of a seperate tool-chain. magnum.np provides a very flexible interface which allows to combine many of these tasks into a single framework. It should bridge the gap between development codes, which are used for the testing of new methods, and production codes which are highly optimized for one specific task.

Complex algorithms can be easily built on top of the available core functions. Possible examples include an eigenmode solver for the calculation of small magnetization fluctuations, the calculation of the dispersion relation of magnonic devices, or the string-method for the calculation of energy barriers between different energy minima \cite{weinan2007simplified, koraltan2020dependence, hofhuis2020thermally}.

Recently, some inverse micromagnetic problems have been reported \cite{wang2021inverse,papp2021nanoscale,kiechle2022experimental}. Due to the use of PyTorch's autograd method magnum.np is also well suited for such applications.

Magnum.np is open-source under the GPL3 licence and can be found at \url{https://gitlab.com/magnum.np/magnum.np}. Different demo scripts are part of the source code and can be tested online using Google Colab \cite{colab}, without the need for local installations or specialized hardware like GPUs. A list of demos can be found on the project gitlab page \url{https://gitlab.com/magnum.np/magnum.np#documented-demos}

\section{Design}
In contrast to many available micromagnetic codes magnum.np follows a high-level approach for easy readability, maintainance and development.
The Python programming language combined with PyTorch offers a powerful environment, which allows to write high-level code, but still get competitive performance due to proper vectorization.

PyTorch\cite{paszke2019pytorch} has been chosen as backend since it allows transparently switching between CPU and GPU without modification of the code. Also the use of single or double precision arithmethic can be switched easily (e.g. use {\tt torch.set\_default\_dtype}).
Furthermore, it offers a very flexible tensor interface, based on the the Numpy Array API. As a nice benefit of using PyTorch, one can directly use inverse operations via the PyTorch's autograd feature \cite{paszke2017automatic}. Even the utilization of deep neural networks in combination with classical micromagnetics would become feasable \cite{kovacs2022magnetostatics}.

One key philosophy of the magnum.np design is to utilize few well-known libraries in order to delegate work, but keep its own code clean and compact. On the other hand we try to keep the number of dependencies as small as possible, in order to improve maintainability. As an example {\tt pyvista} is used for simple reading or writing VTK files, but also offers many additional capabilities (mesh formats, visualization, etc.).

\begin{figure}[h!]
  \begin{center}
  \includegraphics{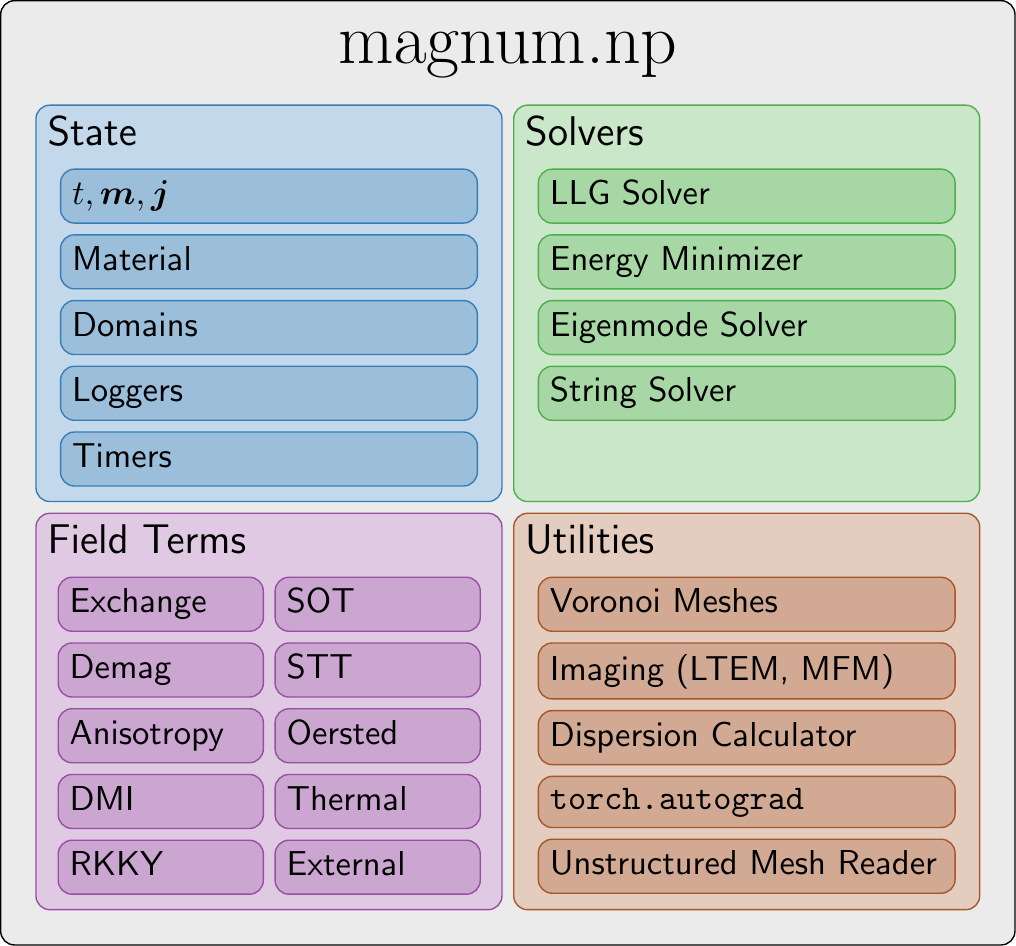}
  \end{center}
  \caption{Overview of the high-level interface of magnum.np.}
  \label{fig:overview}
\end{figure}

Figure~\ref{fig:overview} summarizes the most important building blocks and features.

The {\tt state} class contains the actual state of the simulation like time $t$, magnetization $\vec{m}$ or in case of an Oersted field the corresponding current density $\vec{j}$.
It also contains the information about mesh and materials. 
The finite difference method is based on an equidistant rectangular mesh consisting of $n_x \times n_y \times n_z$ cells, with a grid spacing $(\Delta x, \Delta y, \Delta z)$ and an origin $(x_0, y_0, z_0)$.
Thus the index set $(i,j,k)$ is sufficient to identify an individual cell center:
\begin{align}
\vec{x}_{i,j,k} = \begin{pmatrix} x_i \\ x_j \\ x_k\end{pmatrix} = \begin{pmatrix} x_0 + i \; \Delta x \\ y_0 + j \; \Delta y \\ z_0 + k \; \Delta z \end{pmatrix} = \vec{x}_0 + \Delta \vec{x}
\quad \text{with} \quad \begin{matrix} i = 0...n_x-1 \\ j = 0...n_y-1 \\ k = 0...n_z-1 \end{matrix}
\end{align}
Internally, physical fields are stored as multi-dimensional PyTorch tensors, where one value is stored for each cell (e.g. scalar fields are stored as $(n_x, n_y, n_z, 1)$ tensors).
Using Numpy Array API features like slicing or fancy indexing allows simple modification of the corresponding data.
Furthermore, it allows to use the same expression for constant and non-constant materials, which contains one material parameter for each cell of the mesh. This avoids additional storage in case of constant materials, without the need for independent code branches. By using overloading of the {\tt \_\_call\_\_} operator, it is even possible to allow time dependent material parameters in a transparent way.

Domains are represented by boolean tensors, which can be created by low-level tensor operation or by using {\tt SpatialCoordinate} - a list of tensors $(x,y,z)$ which store the physical location of each cell. Using these coordinate tensors allows to specify domains by simple analytic expressions (e.g. $x^2 + y^2 < r^2$ for a circle with radius $r$). The same coordinate tensors can also be used to parametrize magnetic configurations like vortices or skyrmions (see e.g. Listing~\ref{lst:vortex_parametrization}).
\begin{code}
\begin{minted}[bgcolor=bg]{python}
x, y, z = state.SpatialCoordinate()
r = torch.sqrt(x**2 + y**2)
disk = r < 200e-9

m_vortex  = torch.stack([ y[disk] / r[disk],
                         -x[disk] / r[disk],
                             5e-9 / r[disk]], dim = -1)
\end{minted}
\caption{Parametrization of a vortex configuration of a disk with radius $r = \SI{200}{nm}$ using {\tt SpatialCoordinate}.}
\label{lst:vortex_parametrization}
\end{code}

The actual state can be stored by means of loggers. The {\tt ScalarLogger} is able to log arbitrary scalar functions depending on the current {\tt state} (e.g. average magnetization, field at a certain point, GMR signal, ...). The {\tt FieldLogger} stores arbitrary field data using VTK.

Due to the very flexible interface it is also intendend to add utility function for various application cases to the magnum.np library. In many cases pre- and post-processing is already done in some high-level python scripts, which makes it possible to directly reuse those codes in magnum.np at least on CPU. In many cases time-critical routines can be easily translated into PyTorch code, which then also runs on the GPU, due to the common Numpy Array API. Examples of such utility functions which are already included within magnum.np are Voronoi mesh generators, several imaging tools for post-processing (like LTEM or MFM), or the calculation of a dispersion relation from time-domain micromagnetic simulations.

\section{Landau-Lifshitz-Gilbert Equation}
Dynamic micromagnetism is described by the Landau-Lifshitz-Gilbert Equation
\begin{align}\label{eqn:llg}
\dot{\vec{m}} = -\frac{\gamma}{1+\alpha^2} \left[ \vec{m} \times \vec{h}^\text{eff} + \alpha \, \vec{m} \times \left( \vec{m} \times \vec{h}^\text{eff} \right) \right],
\end{align}
with the reduced magnetization $\vec{m}$, the reduced gyromagnetic ratio $\gamma = \SI{2.21e5}{m/As}$, the dimensionless damping constant $\alpha$, and the effective field $\vec{h}^\text{eff}$.
The effective field may contain several contributions like the magnetostatic strayfield, or the quantummechanical exchange interaction (see Section \ref{sec:field_terms} for the detailed descriptions of possible field terms).

For the solution of the Eqn.~\eqref{eqn:llg} in time-domain most finite difference codes use explicit Runge-Kutta(RK) methods of different order. Magnum.np by default uses the Runge-Kutta-Fehlberg Method (RKF45) \cite{mathews2004numerical}, which uses a 4th order approximation with a 5th order error control. Explicit RK methods, are very common, due to their simplicity and they are well suited for modern GPU computing. Additionally, third party solvers can be easily added, since many libraries already provide a proper python interface. For example wrappers for Scipy(CPU-only) and TorchDiffEq solvers are provided. Those solvers include more complicated solver methods like implicit BDF \cite{suess2002time}, which are well suited for stiff problems.

Often one is only iterested in the magnetic groundstate, in which case the LLG can be integrated with a high damping constant (and optionally without the precession term). Alternatively, the micromagnetic energy \cite{exl2014labonte,exl2019preconditioned} can be minimized directly, which is often much more efficient. However, special care has to be taken since, standard conjugate gradient method may fail to produce correct results \cite{fischbacher2017nonlinear}.

\section{Field Terms}\label{sec:field_terms}
The following section shows some implementation details of the effective field terms. Due to the flexible interface new field terms can easily be added even without modifying the core library.

All field terms which are linear in the magnetization $\vec{m}$ inherit from the {\tt LinearFieldTerm} class, in order to allow a common calculation of the energy using
\begin{align}\label{eq:linear_energy}
\mathcal{E}^\text{lin} = -\frac{1}{2} \mu_0 \int  M_s \, \vec{m} \cdot \vec{h}^\text{lin} \dx,
\end{align}
where $\vec{h}^\text{lin}$ is the corresponding (continuous) field.

In the following several field contributions will be described including a continuous formulation as well as the used discretization.
For example the discretized version of the linear field energy can be written as
\begin{align}\label{eq:linear_energy_discrete}
E^\text{lin} = -\frac{1}{2} \mu_0 V \sum_{\vec{i}} M_s \, \vec{m}_{\vec{i}} \cdot \vec{h}_{\vec{i}}^\text{lin},
\end{align}
with the cell volume $V = \Delta x \, \Delta y \, \Delta z$. $x_{\vec{i}}$ describes a discretized quantity $x$ at the cell with index $\vec{i}$. Some indices $\vec{i}$ will be omitted for sake of better readability (e.g. for the material parameter $M_s$).

\subsection{Anisotropy Field}
Spin orbit coupling gives rise to an anisotropy field which favors the alignment of the magnetization into certain axes. Depending on the crystal structure one or more of such easy axis may be observed. E.g. material with tetragonal or hexagonal structure show a uniaxial anisotropy which gives rise the the following interaction field
\begin{align}
\vec{h}^\text{u}(\vec{x}) = \frac{2 K_\text{u1}}{\mu_0 \, M_s} \; \vec{e}_\text{u} \; (\vec{e}_\text{u} \cdot \vec{m}) + \frac{4 K_\text{u2}}{\mu_0 \, M_s} \; \vec{e}_\text{u} \; (\vec{e}_\text{u} \cdot \vec{m})^3,
\end{align}
where $K_\text{u1}$ and $K_\text{u2}$ are the first and second order uniaxial anisotropy constants, respectively, and $\vec{e}_\text{u}$ is the corresponding easy axis. Since the anisotropy is a local interaction, its discretization is straight forward and will be ommited. The corresponding source code is shown in Listing~\ref{lst:uniaxial_anisotropy}.

For a cubic crystal structure the corresponding cubic anisotropy field is given by
\begin{align}
\vec{h}^\text{c}(\vec{x}) = -\frac{2 K_\text{c1}}{\mu_0 \, M_s} \; \begin{pmatrix} m_1 \, m_2^2 + m_1 \, m_3^2 \\ m_2 \, m_3^2 + m_2 \, m_1^2 \\ m_3 \, m_1^2 + m_3 \, m_2^2\end{pmatrix}
                            -\frac{2 K_\text{c2}}{\mu_0 \, M_s} \; \begin{pmatrix} m_1 \, m_2^2 \, m_3^2 \\ m_1^2 \, m_2 \, m_3^2 \\ m_1^2 \, m_2^2 \, m_3\end{pmatrix},
\end{align}
where $K_\text{u1}$ and $K_\text{u2}$ are the corresponding first and second order cubic anisotropy constants. $m_1$, $m_2$ and $m_3$ are the magnetization components in three orthogonal principal axes.

\begin{code}
\begin{minted}[bgcolor=bg]{python}
class UniaxialAnisotropyField(LinearFieldTerm):
  @timedmethod
  def h(self, state):
    Ku = state.material["Ku"]
    Ku_axis = state.material["Ku_axis"]

    h = 2.*Ku*Ku_axis / (constants.mu_0 * state.material["Ms"]) \
      * torch.sum(Ku_axis * state.m, dim=3, keepdim=True)
    return torch.nan_to_num(h, posinf=0, neginf=0)
\end{minted}
\caption{Implementation of the first order uniaxial anisotropy field.}
\label{lst:uniaxial_anisotropy}
\end{code}

\subsection{Exchange Field}
The quantum mechanical exchange interaction favours the parallel alignment of neigboring spins.
Variation of the micromagnetic energy gives rise the the following exchange field
\begin{align}
\vec{h}^\text{ex}(\vec{x}) = \frac{2}{\mu_0 \, M_s} \nabla \cdot \left( A \, \nabla \vec{m} \right),
\end{align}
combined with a proper boundary condition \cite{abert2019micromagnetics} for the magnetization $\vec{m}$, which can be expressed as
\begin{align}
B = 2 A \, \frac{\partial \vec{m}}{\partial \vec{n}}
\end{align}
The boundary condition is important for the correct treatment of the outer system boundaries, but also for interface between different materials. In general the jump of $B$ over an interface $\Gamma$ needs to vanish ($\jump{B}_\Gamma = 0$). In case of an outer boundary this leads to the well-known $\frac{\partial \vec{m}}{\partial \vec{n}} = 0$, if no further field contibutions (like e.g. DMI) are considered.

The discretized expression of the exchange field considering spacially varying material parameters \cite{heistracher2022proposal} is finally given by
\begin{align}
\vec{h}^\text{ex}_{\vec{i}} = \frac{2}{\mu_0 \, M_{s,\vec{i}}} \; \sum_{k=\pm x, \pm y,\pm z} \frac{2}{\Delta_k^2} \frac{A_{\vec{i}+\vec{e}_k} \; A_{\vec{i}}}{A_{\vec{i}+\vec{e}_k} + A_{\vec{i}}} \; \left( \vec{m}_{\vec{i}+\vec{e}_k} - \vec{m}_{\vec{i}} \right)
\end{align}
where $A$ is the exchange constant and $\Delta_k$ is the grid-spacing in direction $k$. The index $\vec{i} = (i,j,k)$ indicates the cell for which the field should be evaluated, whereas the index $\vec{i} \pm \vec{e}_k$ means the index of the next neighbor in the direction $\pm \vec{e}_k$. Note that the harmonic mean of the exchange constants occurs in front of each next-neighbor difference, which makes it vanish if a cell is located on the boundary. This is important to fulfill the correct boundary conditions $\frac{\partial \vec{m}}{\partial \vec{n}} = 0$.
In case of a homogeneous exchange constant this term simplifies to the well known expression
\begin{align}
\vec{h}^\text{ex}_{\vec{i}} = \frac{2 \, A}{\mu_0 \, M_{s,\vec{i}}} \; \sum_\text{k=x,y,z} \frac{\vec{m}_{\vec{i}+\vec{e}_k} -2 \, \vec{m}_{\vec{i}} + \vec{m}_{\vec{i}-\vec{e}_k}}{\Delta_k^2}
\end{align}

\subsection{DMI Field}
Due to the spin-orbit coupling some materials show an additional antisymmetric exchange interaction called Dzyaloshinskii-Moriya interaction \cite{dzyaloshinsky4thermodynamic,moriya1960anisotropic}. A general DMI field can be written as
\begin{align}\label{eqn:DMI_continuous}
\vec{h}^\text{dmi}(\vec{x}) = \frac{2 \, D}{\mu_0 \, M_s} \; \sum_{k=x,y,z} \vec{e}^\text{dmi}_k \times \frac{\partial \vec{m}}{\partial_k},
\end{align}
with the DMI strength $D$ and the DMI vectors $\vec{e}^\text{dmi}_k$, which describe which components of the gradient of $\vec{m}$ contribute to which component of the corresponding field. It is assumed that $\vec{e}^\text{dmi}_{-k} = -\vec{e}^\text{dmi}_k$.

Different kinds of DMI can be simply implemented by specifying the corresponding DMI vectors. For example the continuous interface DMI field for interface normals in $z$ direction and DMI strength $D_i$ is given by
\begin{align}
\begin{split}
\vec{h}^\text{dmi,i}(\vec{x}) &= -\frac{2 \, D_i}{\mu_0 \, M_s} \; \left[ \nabla \left(\vec{e}_z \cdot \vec{m} \right) - \left(\nabla \cdot \vec{m} \right) \, \vec{e}_z\right] \\
&= \frac{2 \, D_i}{\mu_0 \, M_s} \; \left[\vec{e}_y \times \frac{\partial \vec{m}}{\partial x} - \vec{e}_x \times \frac{\partial \vec{m}}{\partial y}\right],
\end{split}
\end{align}
Thus, the corresponding DMI vectors for interface DMI result in $\vec{e}^\text{dmi} = (\vec{e}_y, -\vec{e}_x, 0)$. See Table \ref{tbl:dmi_types} for a summary of the most common DMI types.

\begin{table}[h!]
  \centering
  \begin{tabular}{|c|c|ll|c|}
    \hline
    DMI  & symmetry & Formula & & DMI Vectors \\
    Type & class    &         & & $\vec{e}^\text{dmi}$ \\
    \hline
    \hline
    Interface     & $C_\text{nv}$ & $\vec{h}^\text{dmi,i}$ & $ = -\frac{2 \, D_i}{\mu_0 \, M_s} \; \left[ \nabla \left(\vec{e}_z \cdot \vec{m} \right) - \left(\nabla \cdot \vec{m} \right) \, \vec{e}_z\right]$ & $(\vec{e}_y, -\vec{e}_x, 0)$ \\
    Bulk          & $T$ or $O$    & $\vec{h}^\text{dmi,b}$ & $ = -\frac{2 \, D_b}{\mu_0 \, M_s} \; \nabla \times \vec{m}$                                                                                        & $(\vec{e}_x, \vec{e}_y, \vec{e}_z)$ \\
    $D_\text{2d}$ & $D_\text{2d}$ &  &                                                                                                                                   & $(-\vec{e}_x, \vec{e}_y, 0)$ \\
    \hline
  \end{tabular}
  \caption{Most common DMI types with the corresponding symmetry class and DMI vectors.}
  \label{tbl:dmi_types}
\end{table}

Finally, equation \eqref{eqn:DMI_continuous} is discretized using central finite differences. For constant $D_i$ this results in
\begin{align}
\vec{h}^\text{dmi}_{\vec{i}} = \frac{2}{\mu_0 \, M_{s,\vec{i}}} \; \sum_{k=\pm x, \pm y,\pm z} \tilde{D}_{\vec{i},k} \; \frac{\vec{e}^\text{dmi}_k \times \vec{m}_{\vec{i}+\vec{e}_k}}{2 \, \Delta_k},
\end{align}
where $\tilde{D}_{\vec{i},k}$ is the effective DMI coupling strength between cell $\vec{i}$ and $\vec{i}+\vec{e}_k$.
Similar to the case of the exchange field, the harmonic mean is used for equal signed coupling strengths, whereas in case of opposing signs the geometric mean of the geometric mean and the arithmetic mean is used:
\begin{align}\label{eqn:D_avg}
\tilde{D}_{\vec{i},k} = \begin{cases}
            \frac{2 \, D_{\vec{i}} \, D_{\vec{i}+\vec{e}_k}}{D_{\vec{i}} \, D_{\vec{i}+\vec{e}_k}}                            & \text{if} \; D_{\vec{i}} \, D_{\vec{i}+\vec{e}_k} \ge 0 \\
            \sign(D_{\vec{i}} \, D_{\vec{i}+\vec{e}_k}) \sqrt{\sqrt{-D_{\vec{i}} \, D_{\vec{i}+\vec{e}_k}} \; |D_{\vec{i}}+D_{\vec{i}+\vec{e}_k}|/2}    & \text{otherwise}
            \end{cases}
\end{align}

Note, that if DMI interactions are in place $\frac{\partial \vec{m}}{\partial \vec{n}} = 0$ does no longer hold. Instead, inhomogeneous Neumann boundary conditions occur (see e.g. equations 11-15 in \cite{vansteenkiste2014design}), which leads to a coupling of exchange and DMI interaction. The exchange field could no longer be calculated independent of the DMI interaction. An alternative formulation is to simply ignore the non-existing values on the boundary, which is consistent with the effective coupling strengths in Eqn.~\eqref{eqn:D_avg}. Although, this approach seems less profound, it has been used in some well-known micromagnetic simulation packages, like {\it fidimag}\cite{bisotti2020fidimag} or {\it mumax3}(openBC)\cite{vansteenkiste2014design}, and shows good agreements for many standard problems \cite{cortes2018proposal}.

\subsection{Demagnetization Field}
\begin{figure}[h!]
  \centering
  \includegraphics{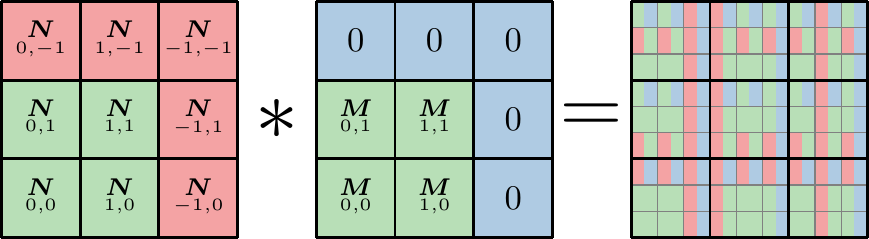}
  \caption{Discrete convolution of the magnetization $\vec{M}$ with the demagnetization kernel $\vec{N}$. The color blocks in the result matrix represent the multiplications of the respective input values. Figure taken from \cite{abert2019micromagnetics} with kind permission of The European Physical Journal (EPJ).}
  \label{fig:convolution}
\end{figure}

\begin{figure}[h!]
  \begin{centering}
  \includegraphics{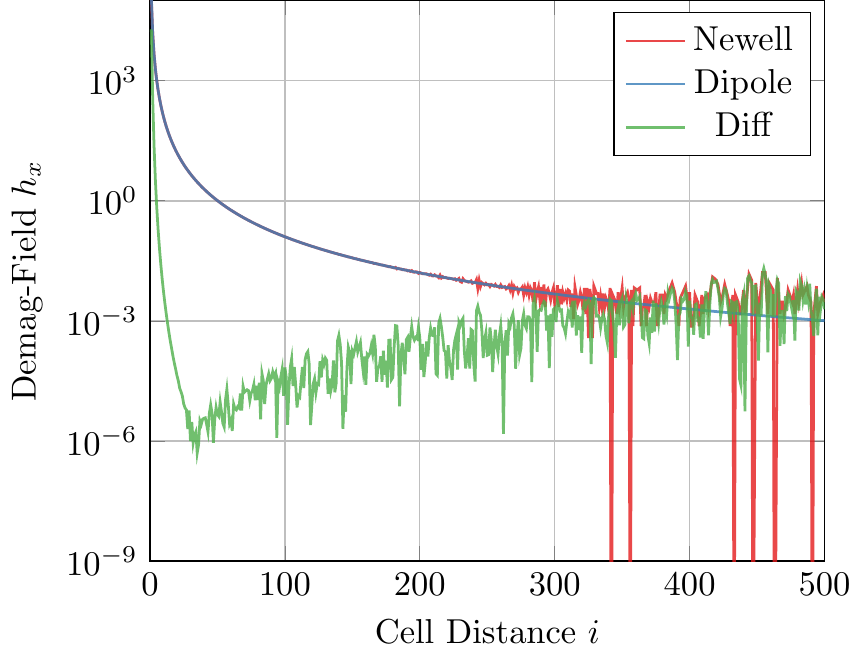}
  \end{centering}
  \caption{Comparison of the numerical strayfield caluclation using Newell's equations \cite{newell1993generalization}, the Dipole appoximation \eqref{eqn:dipole}, and the differnce of both increasing cell distance.}
  \label{fig:demag_longrange}
\end{figure}

The dipole-dipole interaction gives rise to a long-range interaction. The integral formulation of the corresponding Maxwell equations can be represented as convolution of the magnetization with a proper demagnetization kernel $\vec{N}$
\begin{align}\label{eqn:convolution}
\vec{h}^\text{dem}(\vec{x}) = \int\limits_\Omega \vec{N}(\vec{x} - \vec{x}') \, \vec{M}(\vec{x}') \, \dx'
\end{align}
Discretization on equidistant grids results in a discrete convolution which can be efficiently solved by a Fourier method. The discrete convolution theorem combined with zero-padding of the magnetization allows to replace the convolution in real space, with a point-wise multiplication in Fourier space. The discrete version of Eqn.~\ref{eqn:convolution} reads like
\begin{align}
\vec{h}^\text{dem}_{\vec{i}} = \sum\limits_{\vec{j}} \vec{N}_{\vec{i} - \vec{j}} \, \vec{M}_{\vec{j}},
\end{align}
and is visualized in Fig.~\ref{fig:convolution}
The average interaction from one cell to another can be calculated analytically using Newell's formula \cite{newell1993generalization}.
As shown in Fig.~\ref{fig:demag_longrange} the Newell formula is prone to fluctuations if the distance of source and target cell is too large \cite{kruger2013fast}.
Thus, it is favourable to use Newell's formula only for the $p$ next neighbors of a cell. For the long-range interaction one uses a simple dipole field 
\begin{align}\label{eqn:dipole}
  \vec{h}^\text{dipole}(\vec{x}) = \frac{1}{4 \pi} \frac{3 \vec{x} \, (\vec{M} \cdot \vec{x}) - \vert x \vert^2 \, \vec{M}}{\vert x \vert^5},
\end{align}
with the magnetic moment $\vec{M} = V \, M_s \, \vec{m}$ for a cell volume $V$.

The difference of Newell- and dipole-field is also visualized in Fig.~\ref{fig:demag_longrange}. Choosing $p=20$ as default gives accurate results for the near-field,
but avoid fluctuations to the long-range interactions. One further positive effect of using the dipole field for long-range interaction is that the setup of the
demagnetization gets much faster and there is no need for caching the kernel to disk.

In case of multiple thin layers, which are not equi-distantly spaced, it is possible to only use the convolution theorem in the two lateral dimensions \cite{heistracher2020hybrid}. The asymptotic runtime in this case amounts to $\mathcal{O}(n_{xy} \, \log n_{xy} \, n_z^2)$, where $n_{xy}$ are the number of cells within the lateral dimensions and $n_z$ is the number of non-equidistant layers.

True periodic boundary conditions can be used to suppress the influence of the shape anisotropy due to the global demagnetization factor. This is crucial when simulating the microstructure of magnetic materials. The differential version of the corresponding Maxwell equations can be solved efficiently by means of the Fast Fourier Transfrom, which intrinsically fulfills the proper periodic boundary conditions \cite{bruckner2021strayfield}.

\subsection{Oersted Field}
For many applications like the optimization of spinwave excitation antennas \cite{demidov2009excitation, chumak2022advances} or spin orbit torque enabled devices \cite{talmelli2018spin, woo2017spin} the Oersted field created by a given current density has an important influence. For continuous current density $\vec{j}$ it can be calculated by means of the Biot-Savart law
\begin{align}
\vec{h}^\text{oersted}(\vec{x}) = \frac{1}{4 \pi} \int \vec{j}(\vec{x}') \times \frac{\vec{x}-\vec{x}'}{\vert \vec{x}-\vec{x}'\vert^3} \, \dx'
\end{align}

Most common finite difference micromagnetic codes offer the possibility to use arbitrary external fields, but lack the ability to calculate the Oersted field directly. Fortunately, the Oersted field has a similar structure to the demagnetization field and the occuring integral equations can be solved analytically \cite{kruger2011current}. This makes it possible to consider current densities which vary in space and time, since the corresponding field can be updated at each time-step.

As with the demagnetization field the far-field is approximated by the field of a singular current density, which avoids numerical fluctuations.

\subsection{Spin-Torque Fields}
Modern spintronic devices are based on different kinds of spin-torque fields \cite{garello2013symmetry, avci2017current}, which describe the interaction of the magnetization with the electron spin. An overview about models and numerical methods used to simulate spintronic devices can be found in \cite{abert2019micromagnetics}.

In general arbitrary spin torque contributions can be described by the following field
\begin{align}
\vec{h}^\text{st}(\vec{x}) = -\frac{j_e \hbar}{2 e \mu_0 M_s} \left[\eta_\text{damp} \, \vec{m} \times \vec{p} + \eta_\text{field} \, \vec{p} \right],
\end{align}
with the current density $j_e$, the reduced Planck constant $\hbar$, the elementary charge $e$, and the polarization of the electrons $\vec{p}$.
$\eta_\text{damp}$ and $\eta_\text{field}$ are material parameters which describe the amplitude of damping- and field-like torque \cite{abert2017fieldlike}.

In case of Spin-Orbit-Torqe (SOT) $\eta_\text{field}$ and $\eta_\text{damp}$ are constant material parameters,
whereas for the Spin-Transfer-Torque inside of magnetic multilayer structures those parameters additionally depend on $\vartheta$ - the angle between $\vec{m}$ and $\vec{p}$.
Expressions for the angular dependence are e.g. introduced in the original work of Slonczewski \cite{slonczewski2002currents} or more generally in \cite{xiao2005macrospin}.

Spin-Transfer-Torque can also occur in bulk material inside regions with high magnetization gradients like domain walls, or vortex-like structures.
The following field has been proposed by Zhang and Li \cite{zhang2004roles} for this case:
\begin{align}
\vec{h}^\text{stt,zl}(\vec{x}) = \frac{b}{\gamma} \left[\vec{m} \times (\vec{j}_e \cdot \nabla) \vec{m} + \xi \; (\vec{j}_e \cdot \nabla) \vec{m} \right],
\end{align}
with the reduced gyromagnetic ratio $\gamma$, the degree of nonadiabacity $\xi$. $b$ is the polarization rate of the conducting electrons and can be written as
\begin{align}
b = \frac{\beta \mu_B}{e M_s (1+\xi^2)},
\end{align}
with the Bohr magneton $\mu_B$, and the dimensionless polarization rate $\beta$.

The muMAG Standard Problem \#5 is included in the magnum.np source code for demonstration of the Zhang-Li spin-torque.

\subsection{Interlayer-Exchange Field}
The Ruderman-Kittel-Kasuya-Yosida (RKKY) interaction \cite{ruderman1954indirect} gives rise to an exchange coupling of the magnetic layers in multilayer structures which are separated by a non-magnetic layer.
The corresponding continuous interaction energy can be written as
\begin{align}\label{eqn:E_rkky}
E^\text{rkky} = -\int\limits_\Gamma J_\text{rkky} \, \vec{m}_1 \cdot \vec{m}_2 \, \dA,
\end{align}
where $\Gamma$ is the interface between two layers with magnetizations $\vec{m}_1$ and $\vec{m}_2$, respectively.
$J_\text{rkky}$ is the coupling constant which oscillates with respect to the spacer layer thickness.

When discretizing the RKKY field using finite difference in many cases the spacer layer is not discretized.
Instead the interaction constant $J_\text{rkky}$ is scaled by the spacer layer thickness.
Additionally, one has to make sure that the two layers are not coupled by the classical exchange interaction.
In magnum.np the corresponding exchange field can be defined on subdomains, so there is no coupling via the interface.

The magnetizations $\vec{m}_1$, $\vec{m}_2$ should be evaluated directly at the interface.
Since the magnetization is only available at the cell centers, most finite difference codes use a lowest order approximation which directly uses those center values.  magnum.np also allows to use higher order approximations, which show significantly better convergence if partial domain walls are formed at the interface \cite{suess2022accurate}.

For $\vec{m}_1$ the following expression can be found:
\begin{align}
  \vec{m}_1 =
    \begin{cases}
      \vec{m}_{\vec{i}} & \text{if order} = 0 \\
      \frac{3}{2} \, \vec{m}_{\vec{i}} - \frac{1}{2} \, \vec{m}_{\vec{i}-1} & \text{if order} = 1 \\
      \frac{15}{8} \, \vec{m}_{\vec{i}} - \frac{5}{4} \, \vec{m}_{\vec{i}-1} + \frac{3}{8} \, \vec{m}_{\vec{i}-2} & \text{if order} = 2 \\
    \end{cases}
\end{align}
where $\vec{m}_{\vec{i}}$ denotes the magnetization of the cell adjacent to the interface insided of layer 1, where the field should be evaluated. $\vec{m}_{\vec{i}-1}$ and $\vec{m}_{\vec{i}-2}$ are its first and second next neighbor, respectively. A similar expression is given for $\vec{m}_2$, but indices $\vec{i}$ are replaced with the corresponding indices $\vec{j}$ of cells inside of layer 2.

Finally, the discretization of the RKKY field corresponding to the energy Eqn.~\eqref{eqn:E_rkky} yields
\begin{align}
\vec{h}^\text{rkky}_{\vec{i}} = \frac{J_\text{rkky}}{\mu_0 \, M_s \, \Delta_z} \left[\vec{m}_2 - \left(\vec{m}_1 \cdot \vec{m}_2 \right) \vec{m}_1 \right],
\end{align}
with the cell thickness $\Delta_z$ and the indices $\vec{i}$ and $\vec{j}$ of two adjacent cells in layer $i$ and $j$. Note that the second term stems from a modified boundary condition for the classical exchange field, if higher order approximations are used.

\subsection{Thermal Field}
Thermal fluctuation can be considered in micromagnetic simulations by adding a stochastic thermal field $\vec{h}^\text{th}$, which is characterized by
\begin{align}
\begin{split}
\langle \vec{h}^\text{th}_{\vec{i}}\rangle &= 0 \\
\langle \vec{h}_{\vec{i}}^\text{th}(t_0) \; \vec{h}_{\vec{j}}^\text{th}(t_1)\rangle &= \frac{2 \alpha k_B T}{\mu_0 M_s \gamma V \Delta t} \; \delta(t_1-t_0) \; \delta_{\vec{i}\vec{j}}
\end{split}
\end{align}
with the Boltzmann constant $k_B$, the temperature $T$, the dimensionless damping parameter $\alpha$, the cell volume $V$, and the timestep $\Delta t$. $\langle . \rangle$ denotes the ensemble average. The two delta functions indicate that the thermal noise is spatially and temporally uncorrelated. 
The actual thermal field can then be calculated by
\begin{align}
\vec{h}^\text{th}_{\vec{i}} = \vec{\eta}_{\vec{i}} \sqrt{\frac{2 \alpha k_B T}{\mu_0 M_s \gamma V \Delta t}},
\end{align}
where $\vec{\eta}_{\vec{i}}$ is a random vector drawn from a standard normal distribution for each time-step.

When numerically integrating stochastic differential equations, a drift term can occur if not using the correct statistics within the numerical methods. Although some higher-order Runge-Kutta schemes exist, they become increasingly complex. Fortunately, it has been proven that in case of the LLG the drift term only changes the length of the magnetization, which is fixed anyway. Thus, it is possible to straight forwardly use available adaptive higher order schemes for the solution of the stochastic LLG \cite{leliaert2017adaptively}.

\subsection{Timings}
\begin{figure}[h!]
  \begin{centering}
    \includegraphics{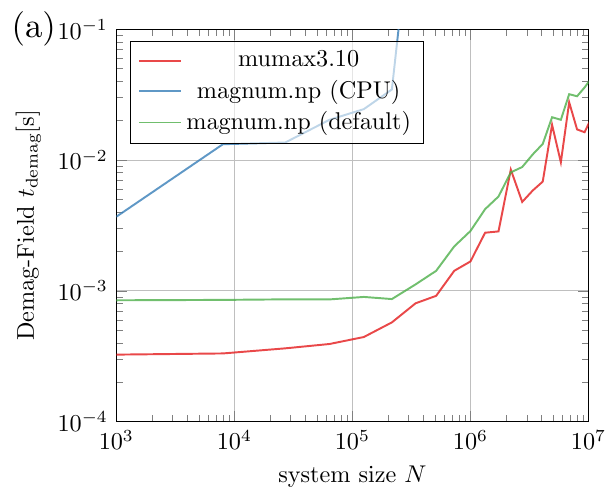}
    \includegraphics{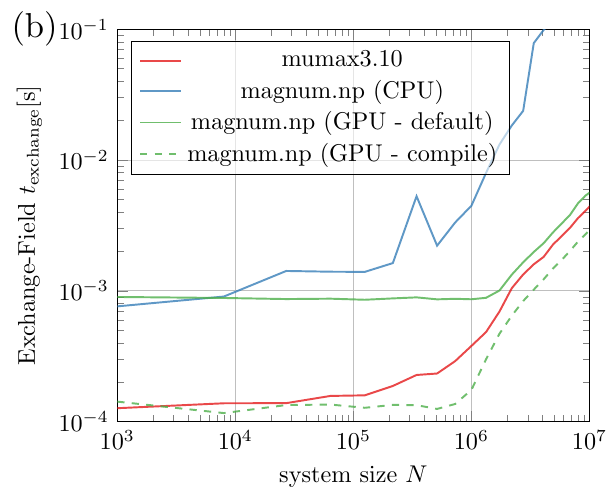}
  \end{centering}
  \caption{Benchmarking (a) demagnetization field and (b) exchange field for different system sizes $N$ on an Intel(R) Xeon 6326 CPU @ 2.90GHz using one NVIDIA A100 80GB GPU (CUDA Driver 11.8). An average of 10000 evaluations has been measured for each field term. Before measurent begins, 1000 warm-up loops are used to ensure that the GPU has reached its maximum performance state. Single precision arithmetics are used for comparison with mumax3.
}
  \label{fig:timings}
\end{figure}

Benchmarks of the field terms are presented in Figure~\ref{fig:timings}. The results show that for systems larger than about $N = 10^6$ elements, the demagnetization field is the dominating field term and it is less than a factor 2 slower than the mumax3 version. However, these timings have been performed without any low-level optimization. Instead magnum.np utilizes high-level optimization, that does not influence the simplicity of the code. For example just-in-time compilers (like PyTorch-compile, numba, nvidia-warp, etc.) are used to improve the performance of the code. For all local field contributions this works increadibly well and the resulting timings are even outperforming mumax3. Optimized timing using {\tt torch.compile} of the recently published version 2.0 of PyTorch are included in Figure~\ref{fig:timings}. 
Unfortunately, {\tt torch.compile} does not yet support complex datatypes, which prevents it from being used to calculate the demagnetization field.

In case of the demagnetization field an optimized padding for the 3D FFT which is not yet provided by PyTorch, could give some further speedup.

\section{Examples}
The following section provides some examples which should demonstrate the ease of use and the power of the magnum.np interface. Due to the python / PyTorch interface pre- and post-processing can be done in a single script (or at least in the same scripting language) and allows to keep the complete simulation framework as simple as possible. The presented code focuses on complex examples which would be more elaborate to setup with other micromagnetic codes. 
In the magnum.np source code \cite{gitlab_magnumnp} several other examples are included, such as hysterses loop calculations, simulation of soft magnetic composites, an RKKY standard problem and the muMAG standard problems. Further examples will be continously added. 

\subsection{Spintronic Devices}
The first example demonstrates the creation and manipulation of skyrmions in magnetic thin films, that can be patterned by means of ion radiation techniques to locally alter the magnetic materials of the system\cite{kern2022deterministic}. This simulation technique is also useful for the numerical modeling of structued Pt-layers on top of the thin-film that create a location-dependent DMI interaction as realized recently in an experimental work \cite{velez2022current}.

Listing~\ref{lst:setup} shows the material definition for the spintronic demo, where the anisotropy constant is altered in the irradiated region. 
A rectangular mesh with $\vec{n}$ cells and a grid spacing $\vec{dx}$ is created and domains are read from an unstructured mesh file by means of the {\tt mesh\_reader}.
Those domain ids can then be used to set location dependent material parameters, which in turn influences the local skyrmion densities.

\begin{code}
\begin{minted}[bgcolor=bg]{python}
mesh = Mesh(n = (1500, 300, 1), dx = (10e-9, 10e-9, 20e-9))
domains = read_mesh(mesh, "mesh.msh", scale = 1e-9)
state = State(mesh)

irradiated = (domains == 1)
asdeposited = (domains == 2)
magnetic = (irradiated | asdeposited)

state.material = {
    "Ms": 400e3,
    "A": 4e-12,
    "Ku_axis": (0,0,1),
    "alpha": 1.0}

state.material["Ku"] = state.Constant([100e3])
state.material["Ku"][irradiated] = 50e3

write_vti(state.material, "material.vti")
\end{minted}
\caption{Mesh creation and boolean domains read from an external unstructured grid file "mesh.msh", which are used to define location dependent material parameters.}
\label{lst:setup}
\end{code}

\begin{figure}[h!]
  \centering
  \includegraphics{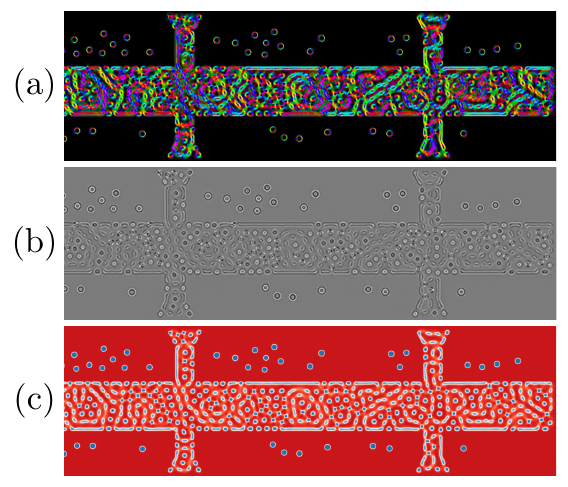}
  \caption{Visualization of the created skyrmions at $\mu_0 H_z=\SI{250}{mT}$ using (a) an Induction Map, (b) an underfocus Lorentz Transmission Electron Microscopy image, or (c) the $z$-component of the magnetization.}
  \label{fig:skyrmion_demo}
\end{figure}

A random initial magnetization is set and the default RKF45 solver is used for time-integration. Several logging capabilities allow to flexibly log scalar- and field-data to files.
Custom python functions that return derived quantities, such as the Induction Map (IM) or the Lorentz Transmission Electron Microscopy (LTEM) image of the magnetization state, can simply be added as log entries. 
Listing~\ref{lst:llg} shows the corresponding code and the results are visualized in Fig.~\ref{fig:skyrmion_demo}.
One can see that the density of skyrmions in the irradiated region is increased significantly compared to the outside region. The lower anisotropy allows the nucleation of not only skyrmions, but also trivial type-II bubbles, and antiskyrmions\cite{heigl2021dipolar}.

\begin{code}
\begin{minted}[bgcolor=bg]{python}
state.m = RandomUnitSphere(state)

demag    = DemagField()
aniso    = UniaxialAnisotropyField()
exchange = ExchangeField()
external = ExternalField(Hext)

llg = LLGSolver([demag, aniso, exchange, external])
logIM = ('IM', lambda state: IMImage(state))
logLTEM = ('LTEM', lambda state: LTEMImage(state))

logger = Logger("data", ['t','m'], ['m', logIM, logLTEM])
while state.t < 20e-9:
    logger << state
    llg.step(state, 1e-11)
\end{minted}
\caption{Setup of time-integration for \SI{20}{ns} and logging. Scalar data, like time $t$ and avarage magnetization $\langle \vec{m} \rangle$, will be written to a column based text field. Field data, like the magnetization $\vec{m}$ as well as a corresponding LTEM image, will be written to {\tt .vti} files utilizing {\tt pyvista}.}
\label{lst:llg}
\end{code}

\subsection{Inverse Design}
Finding the optimal shape of magnetic components for certain applications is an essential, but quite challenging task. An automated topology optimization requires the efficient calculation of the so-called forward problem, as well as the corresponding gradients (compare e.g. \cite{abert2017fast, huber2017topology}). The following example should demonstrate how magnum.np can be used to solve inverse problems, by utilizing PyTorch's autograd mechanism.

The field created by a magnetization at a certain location $\vec{x}_0$ should be maximized (e.g. $J[\vec{m}] = h_y(\vec{x}_0)$). Thus, the forward problem is simply an evaluation of the demagnetization field. The optimization requires the calculation of the gradient $\vec{g} = \frac{\partial J}{\partial \vec{m}}$. The magnetization should always point in $y$ direction, and its magnitude $m_y$ saturates at $M_s$.

\begin{figure}[h!]
  \centering
  \includegraphics{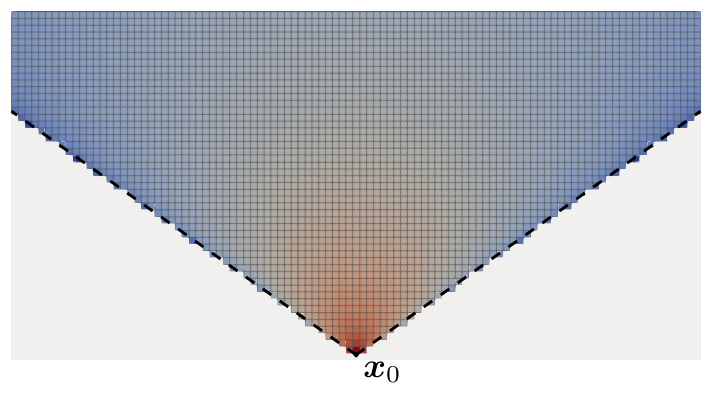}
  \caption{Optimal topology that maximizes the z-component of the strayfield at the marked cell. Only cells with a positive gradient are shown. The logarithmic color scheme represents the sensitivity of the objective function on the magnetization within the corresponding cell (red means a large sensitivity). The dotted line shows the analytic result $x < \sqrt{2} \, y$.}
  \label{fig:inverse_demo}
\end{figure}

Since this simple example is linear, the optimal solution is found after a single iteration. Depending on the sign of the gradient. The optimal magnetization within each cell is 1, if the calculated gradient is positive and 0 otherwise. Listing~\ref{lst:inverse_demo} summarizes how the gradient calculation is performed. The optimized magnetization is visualized in Fig.~\ref{fig:inverse_demo} and shows perfect agreement with the analytical result.

\begin{code}
\begin{minted}[bgcolor=bg]{python}
from magnumnp import *

n  = (101, 51, 1)
dx = (5e-9, 5e-9, 5e-9)
mesh = Mesh(n, dx)
state = State(mesh)
state.material = {"Ms": 1.}

state.m = state.Constant([0,1,0], requires_grad = True)
h = DemagField().h(state)
J = h[n[0]//2, 0, n[2]//2, 1]

J.backward()
write_vti(state.m.grad, "data/m_grad.vti")
\end{minted}
\caption{Full topology optimization example which solves the inverse strayfield problem utilizing PyTorch's autograd mechanism.}
\label{lst:inverse_demo}
\end{code}

\section{Conclusion}
An overview of the basic design ideas of magnum.np has been given. Equations and references for the most important field contributions as well as solving methods are included for clarification. Some typical applications are provided in order to demonstrate the ease of use and the power of the provided python-base interface. Furthermore the use of PyTorch extends magnum.np's capabilities to inverse probems and allows seamlessly running applications on CPU and GPU without any modification of the code. The openness of the project should encourage other developers to contribute code and use magnum.np as a framework for the development and testing of new algorithms, while still getting reasonable performance and generality.

\appendix

\section{Acknowledgments}
This research was funded in whole, or in part, by the Austrian Science Fund (FWF) P 34671 and (FWF) I 4917. For the purpose of open access, the author has applied a CC BY public copyright licence to any Author Accepted Manuscript version arising from this submission.

\bibliographystyle{ieeetr}
\bibliography{refs}

\end{document}